# Application of Transformation Optics for the Purpose of Cloaking

**Abstract**

The advent of transformation optics has lead to the initiation of designing devices and applications associated to electromagnetic wave propagation in anisotropic media. Here, a method is suggested using a coordinate transformation with spherical coordinates for the purpose of designing a three dimensional cloak for a body having an arbitrary convex geometry. For the purpose of verification of the algorithm, a ray tracing process is carried out for an ellipsoid having axial symmetry.

**Sidhwa HH***

*Department of Electrical Engineering, Indian Institute of Technology Bombay, India*

**\*Corresponding author:** Haroonhaider Sidhwa, Department of Electrical Engineering, Indian Institute of Technology Bombay, India, Email: haroonhaider.sidhwa@gmail.com

## Introduction

Transformation optics is an allied area comprising of differential geometry and electrodynamics [1]. The quest of humans to attain invisibility seems credible with the realisation of this field. The conceptualization of invisibility has been indicated in scientific parlance as cloaking which can be explained as bending the path of the ray around the body of interest in order to masquerade its path undistorted by any encumbrance [2]. Pendry et al. [3] and Leonhardt [4] commenced the idea of cloaking in 2006 by carrying out a coordinate transformation in order to alter the material characteristics, which would in turn cause a change in the formulation of Maxwell equations. A process of ray tracing in transformed media for spherical and cylindrical cloaks using Cartesian tensors was carried out by Pendry et al. [5]. A generalised method for designing arbitrarily shaped cloaks using the approach of coordinate transformation has been discussed by C Li & F Li [6]. A full wave simulation using a commercial software program was carried out by them for its verification. Other techniques for cloaking include a scattering cancellation method wherein scattering can be minimized by covering the main object by a single layer or by multiple layers of dielectric materials [7,8]. Another technique based on the usage of volumetric structures composed of two dimensional or three dimensional transmission line networks, is illustrated [9]. In this paper, we propose an algorithm for cloaking of an arbitrarily shaped body in three dimensions. This generalised transformation technique would enable designing of a cloak with an arbitrary geometry without the need to calculate the material characteristics explicitly but which can easily be found with the help of Jacobian as explained later. The Hamiltonian is calculated using a technique similar to that reported in [5] but which obviates the need to calculate the determinant of refractive index, which is cubersome in nature. For the purpose of verification of the algorithm, it is assumed that the arbitrary body is an ellipsoid having axial symmetry.

## Transformation of Coordinate System

Consider an arbitrarily shaped body to be cloaked which is enclosed by another body with the same topology. A coordinate transformation is carried out to map any point lying within the region from the centre to the boundary of the outer body to the region lying between the two bodies.

Consider $O$ to be the centre of the concentric bodies as shown in Figure 1. Let $A$ be any point on the surface of the inner body and $A'$ be the corresponding point having the same value of ($\theta, \varphi$) on the surface of the outer body in the radial direction. The transformation is carried out such that any point lying in the region $OA$ gets transformed to a point lying in the region $AA'$.

$$\frac{OA}{OA'} = \tau \qquad (1)$$

$$\mathbf{r}' = R_0(\theta, \varphi)\hat{r} \qquad (2)$$

$$\mathbf{r} = \tau R_0(\theta, \varphi)\hat{r} \qquad (3)$$

Where $\hat{r}$ is the unit vector in the radial direction, $\mathbf{r}$ is the position vector of a point on the surface of the body to be cloaked for a given ($\theta, \varphi$), $\mathbf{r}'$ is the position vector of the point on the surface of the cloak for the same ($\theta, \varphi$), and $R_0$ is a mathematical function that defines the nature of contour of the surface of the cloaked and outer bodies which is dependent on ($\theta, \varphi$). Since the cloak parameters depend on ($\theta, \varphi$), for the unique definition of the parameters, the function $R_0$ must be a single valued function of ($\theta, \varphi$).

$$r' = (1-\tau)r + \tau R_0 \qquad (4)$$

$$\theta' = \theta \quad \varphi' = \varphi \qquad (5)$$

$$r' = \tau R_0 ; \quad \text{if} \quad r = 0 \qquad (6)$$

$$= (1-\tau)R_0 + \tau R_0 \quad \text{if} \quad r = R_0 \qquad (7)$$

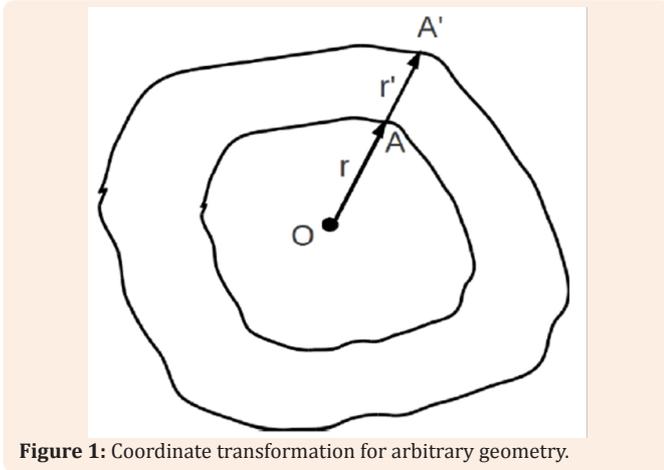

**Figure 1:** Coordinate transformation for arbitrary geometry.

The transformation from unprimed to primed coordinates preserves the direction of the position vector **r**, which means that **r**, **r'** are parallel.

$$\frac{x'}{r'} = \frac{x}{r}; \quad \frac{y'}{r'} = \frac{y}{r}; \quad \frac{z'}{r'} = \frac{z}{r} \qquad (8)$$

Expressing the equation in the tensor notation and using Equation (4),

$$x' = \frac{r'}{r} x \qquad (9)$$

$$x'_i = \left(\tau \frac{R_0}{r} + (1-\tau)\right) x_j \delta_{ij} \qquad (10)$$

$$x'_i = (1-\tau) x_j \delta_{ij} + \tau R_0 \frac{x_j}{r} \delta_{ij} \qquad (11)$$

Where $\delta_{ij}$ is the Kronecker delta function.

The Jacobian for the coordinate transformation can be written as shown in [5]:

$$\Lambda_l^{i'} = \frac{\partial x'_i}{\partial x_l} = (1-\tau)\frac{\partial x_j}{\partial x_l}\delta_{ij} + \tau R_0 \frac{\partial}{\partial x_l}\left(\frac{x_j}{r}\right)\delta_{ij}$$
$$+\tau \frac{x_j}{r}\frac{\partial R_0}{\partial x_l}\delta_{ij} = \frac{r'}{r}\delta_{il} - \frac{R_0}{r^3}\tau x_i x_l + \tau \frac{x_i}{r}\left[\frac{\partial R_0}{\partial \theta}\frac{\partial \theta}{\partial x_l} + \frac{\partial R_0}{\partial \varphi}\frac{\partial \varphi}{\partial x_l}\right]. \qquad (12)$$

The gradient of a function F in spherical coordinates can be expressed as

$$\nabla F(r,\theta,\varphi) = \frac{\partial F}{\partial r}\hat{r} + \frac{1}{r}\frac{\partial F}{\partial \theta}\hat{\theta} + \frac{1}{r\sin\theta}\frac{\partial F}{\partial \varphi}\hat{\varphi} \qquad (13)$$

$$\frac{\partial \theta}{\partial x_l} = \frac{\hat{\theta}_l}{r} \quad \frac{\partial \varphi}{\partial x_l} = \frac{\hat{\varphi}_l}{r\sin\theta}. \qquad (14)$$

Equation (14) can be used to express $\Lambda_l^{i'}$ (12) as

$$\frac{\partial x'_i}{\partial x_l} = \frac{r'}{r}\delta_{il} - \frac{\tau R_0}{r^3} x_i x_l + \tau \frac{x_i}{r^2}\left[\frac{\partial R_0}{\partial \theta}\hat{\theta}_l + \frac{\partial R_0}{\partial \varphi}\frac{\hat{\varphi}_l}{\sin\theta}\right]. \qquad (15)$$

## Formulation of Hamiltonian

The media properties ($\mu, \varepsilon$) in the transformed medium can be expressed as:

$$\mu' = \varepsilon' = \frac{\Lambda \Lambda^T}{|\Lambda|} \qquad (16)$$

$$|\mu'| = |\varepsilon'| = \frac{|\Lambda||\Lambda^T|}{(|\Lambda|)^3} = \frac{1}{|\Lambda|}. \qquad (17)$$

The motion of a material particle is determined by the Hamilton-Jacobi equation. The function of a wave vector in geometric optics is same as that of momentum of a particle in mechanics, while the frequency is akin to Hamiltonian, i.e., the energy of the particle.

Since there is no loss of energy while the wave propagates through the medium, the Hamiltonian can be found in order to trace the path of the wave in the cloaked medium [5].

From the classical theory of electromagnetic [10], the dispersion equation in an anisotropic medium can be expressed as:

$$An^4 - Bn^2 + C = 0 \qquad (18)$$

$$A = (\mathbf{k} \cdot \varepsilon' \cdot \mathbf{k})(\mathbf{k} \cdot \mu' \cdot \mathbf{k}) \qquad (19)$$

$$B = \left(\mathbf{k} \cdot \mu' \cdot \left\{[adj(\varepsilon') \cdot \mu']_t \mathbf{I} - adj(\varepsilon') \cdot \mu'\right\} \cdot \mathbf{k}\right) \qquad (20)$$

$$C = |\varepsilon' \cdot \mu'| \qquad (21)$$

Here $n$ is the refractive index of the medium, $I$ is the unity matrix in a three dimensional system, $adj(\varepsilon')$ is the adjoint matrix of $\varepsilon'$ and $[adj(\varepsilon') \cdot \mu']_t$ is the trace of the product of the adjoint of $\varepsilon'$ and $\mu'$. Under the condition that $\varepsilon' = \mu'$, which is the case for cloaking, the above equation reduces to Equation (22) same as Equation (37) in [5] and as explained below.

$$\left(\mathbf{K}\mu^{-1}\mathbf{K} + \varepsilon\right)\mathbf{E_0} = 0 \qquad (22)$$

$$\therefore |\mathbf{K}\mu^{-1}\mathbf{K} + \varepsilon| = 0 \qquad (23)$$

$$\because \mu = \varepsilon = n = \sqrt{\mu\varepsilon}$$

$$|\mathbf{K}n^{-1}\mathbf{K} + n| = \frac{1}{|n|}\left(kn\mathbf{k} - |n|\right)^2 \qquad (24)$$

$$H = \left(kn\mathbf{k} - |n|\right) = 0 \qquad (25)$$

Where $n$ is the refractive index of the medium, $\mathbf{K} = \mathbf{k} \times I$ and $I$ is the unity matrix in a three dimensional system. The Hamiltonian can be expressed as

$$H = <\mathbf{k}|\mu'|\mathbf{k}> - |\mu'| \qquad (26)$$

$$= <\mathbf{k}|\frac{\Lambda\Lambda^T}{|\Lambda|}|\mathbf{k}> - \frac{1}{|\Lambda|}. \qquad (27)$$

Since $det\Lambda$ i.e. $|\Lambda|$ is a function which is not dependent on the Hamiltonian, $|\Lambda|$ can be eliminated. The Hamiltonian would now read:

$$H = <\mathbf{k} | \Lambda\Lambda^T | \mathbf{k}> - 1 = 0 \quad (28)$$

In order to illustrate the similarity of approach between Hamiltonian mechanics and Maxwell equations, we consider the spherical cloak as described by Pendry [5]. One can start with the Maxwell equations, write the dispersion equation Hamiltonian H=0 (Equation 25), take its gradient with respect to $\mathbf{k}'$. For a spherical cloak, the Hamiltonian can be written as [5]:

$$H = <\mathbf{k} | \Lambda\Lambda^T | \mathbf{k}> - \frac{\omega^2}{c^2}\left[\frac{b(r-a)}{r(b-a)}\right]^2 \quad (29)$$

The gradient of the dispersion equation (Equation 29) can be written as [10]:

$$\frac{\partial H}{\partial \mathbf{k}} + \frac{\partial H}{\partial \omega}\frac{\partial \omega}{\partial \mathbf{k}} = 0 \quad (30)$$

$$2|\Lambda\Lambda^T|\mathbf{k}> - \frac{\omega^2}{c^2}\left[\frac{b(r-a)}{r(b-a)}\right]^2 \frac{1}{\omega}\frac{\partial \omega}{\partial \mathbf{k}} = 0 \quad (31)$$

The group velocity can be defined as $\frac{\partial \omega}{\partial \mathbf{k}}$. The ray vector $\mathbf{s}$ can be defined as

$$\mathbf{s} = \frac{1}{k_0^2}\left[\frac{b(r-a)}{r(b-a)}\right]|\Lambda\Lambda^T|\mathbf{k}> \quad (32)$$

The ray vector $\mathbf{s}$ which indicates the direction of flow of power is in the same direction as $\frac{d\mathbf{x}}{d\varsigma}$: Equation (35). As explained in [5,11]. Since the media properties ($\mu, \varepsilon$) in the transformed medium, Equation (16) are changing with position in the problem we are dealing with as well as in the case of Pendry [5], Equation (37) of [5] cannot be solved for any arbitrary position, hence the fields cannot be calculated analytically. One is compelled to use Hamiltonian mechanics (as discussed in [5]) to solve the dispersion relation to find $\mathbf{k}$ as a function of position and free space.

The transformation equation Eq.15 can be expressed in terms of primed coordinates and the ratio ($r'/r$) can be eliminated later.

$$\Lambda_l^{i'} = \frac{r'}{r}\left[\delta_{il} - \frac{x_i'}{r'}\frac{x_l'}{r'}\frac{\tau R_0}{r'} + \tau \frac{x_i'}{r'}\left(\frac{\partial R_0}{\partial \theta}\hat{\theta}_l + \frac{\partial R_0}{\partial \varphi}\frac{\hat{\varphi}_l}{\sin\theta}\right)\frac{1}{r'}\right]. \quad (33)$$

The aim is to express the whole expression in terms of the primed coordinate system. Since the Hamiltonian involves the product $\Lambda\Lambda^T$, the ratio $\left(\frac{r'}{r}\right)^2$ can eliminated completely by multiplying both sides by $\left(\frac{r}{r'}\right)^2$ using Equation (4) as

$$\left(\frac{r}{r'}\right)^2 = \frac{(r' - \tau R_0)^2}{(1-\tau)^2 r'^2}.$$

Further analysis will be carried out in terms of $r'$, but the primes will be dropped.

For the sake of simplicity, we consider an axisymmetric cloak i.e. $\frac{\partial R_0}{\partial \varphi} = 0$. The Hamiltonian is expressed in terms of the transformed coordinate system. On substituting the corresponding values and carrying out a nontrivial derivation, we obtain

$$H = \mathbf{k}\cdot\mathbf{k} + \frac{\left[(\tau R_0)^2 - 2r\tau R_0\right]}{r^4}(\mathbf{k}\cdot\mathbf{x})^2 - \frac{(r - \tau R_0)^2}{(1-\tau)^2 r^2} +$$

$$\frac{\tau^2}{r^4}\left(\frac{\partial R_0}{\partial \theta}\right)^2(\mathbf{k}\cdot\mathbf{x})^2 + \frac{2\tau}{r^2}\frac{\partial R_0}{\partial \theta}(\mathbf{k}\cdot\mathbf{x})(\mathbf{k}\cdot\hat{\theta}) \quad (34)$$

$$-\frac{\tau^2 R_0}{r^5}\frac{\partial R_0}{\partial \theta}(\mathbf{k}\cdot\mathbf{x})^2(\mathbf{x}\cdot\hat{\theta}).$$

For a spherical cloak, only first three terms of Equation 31 exist since

$$\frac{\partial R_0}{\partial \theta} = 0.$$

This reduced equation of Hamiltonian agrees with the expression of Hamiltonian for spherical cloak given in [5]. The transmitted wave vector $\mathbf{k}_t$ must lie in the plane of incidence.

$$\mathbf{k}_t = \mathbf{k}_{in} + q\mathbf{N}$$

where $\mathbf{k}_{in}$ is the incident wave vector, $\mathbf{N}$ is the normal vector pointing inwards to the plane of incidence, and $q$ is a scalar quantity which is obtained by solving for $H = 0$ using a procedure similar to that described by Pendry [5].

For the purpose of ray tracing, the path can be parameterised using the Hamiltonian, Equation 34 as [11]:

$$\frac{d\mathbf{x}}{d\varsigma} = \frac{\partial H}{\partial \mathbf{k}} \quad (35)$$

$$\frac{d\mathbf{k}}{d\varsigma} = -\frac{\partial H}{\partial \mathbf{x}} \quad (36)$$

Where $\varsigma$ is the parameterising variable and $\mathbf{x}$ is the position vector. Ray tracing is carried out by solving $(\mathbf{x}, \mathbf{k})$ as a function of $\varsigma$ starting from $\varsigma = 0$ which corresponds to the point at which the incident wave touches the cloak. In a spherical coordinate system, the position vector $\mathbf{x}$ points in the direction of the radial vector $\hat{r}$, which means $\mathbf{x} = \mathbf{r} = r\hat{r}$.

The Hamiltonian can be solved as per [11] as:

$$\frac{\partial H}{\partial \mathbf{k}} = 2\mathbf{k} + \frac{\left[(\tau R_0)^2 - 2r\tau R_0\right]}{r^4}2(\mathbf{k}\cdot\mathbf{x})\mathbf{x} + 2\frac{\tau^2}{r^4}\left(\frac{\partial R_0}{\partial \theta}\right)^2(\mathbf{k}\cdot\mathbf{x})\mathbf{x}$$

$$+\frac{2\tau}{r^2}\frac{\partial R_0}{\partial \theta}\left[(\mathbf{k}\cdot\mathbf{x})\hat{\theta} + (\mathbf{k}\cdot\hat{\theta})\mathbf{x}\right] \quad (37)$$

$$\frac{\partial H}{\partial \mathbf{x}} = \left[ \left( \frac{6\tau r R_0 - 4(\tau R_0)^2}{r^5} \right)(\mathbf{k}\cdot\mathbf{x})^2 - 2\frac{(r-\tau R_0)\tau R_0}{(1-\tau)^2 r^3} \right.$$

$$+ \left[ \frac{2\tau}{r^5}\left(\frac{\partial R_0}{\partial \theta}\right)(R_0 - r)(\mathbf{k}\cdot\mathbf{x})^2 + 2\frac{(r-\tau)\tau R_0}{(1-\tau)^2 r^3}\left(\frac{\partial R_0}{\partial \theta}\right) \right]$$

$$+ \frac{2\tau^2}{r^5}\left(\frac{\partial R_0}{\partial \theta}\right)\left(\frac{\partial^2 R_0}{\partial \theta^2}\right)(\mathbf{k}\cdot\mathbf{x})^2 - \frac{(\mathbf{k}\cdot\mathbf{r})^2}{r^2}\frac{2\tau}{r^2}\left(\frac{\partial R_0}{\partial \theta}\right)$$

$$+ \frac{2\tau}{r^3}\left(\frac{\partial^2 R_0}{\partial \theta^2}\right)(\mathbf{k}\cdot\mathbf{r})(\mathbf{k}\cdot\hat{\theta})\bigg]\hat{\theta}$$

$$+ \left[ \frac{\left((\tau R_0)^2 - 2r\tau R_0\right)}{r^4}2(\mathbf{k}\cdot\mathbf{r}) + \frac{2\tau^2}{r^4}\left(\frac{\partial R_0}{\partial \theta}\right)^2(\mathbf{k}\cdot\mathbf{r}) \right]$$

$$+ \frac{2\tau}{r^2}\left(\frac{\partial R_0}{\partial \theta}\right)(\mathbf{k}\cdot\hat{\theta})\bigg]\mathbf{k}. \quad (38)$$

In order to calculate the Jacobian, a spherical coordinate system $(\theta, \varphi)$ has been used. However, for the case of ray tracing wherein the calculation of distance is involved, the vectors $\hat{r}, \hat{\theta}, \mathbf{k}$ are expressed in the corresponding Cartesian components.

## Ellipsoidal Cloak

For the verification of the above algorithm, the outer contour is considered to be an ellipsoid having axial symmetry. The ellipsoid would look identical for any section containing the axis of symmetry. Since we are assuming a spherical polar axis of symmetry which is the Z axis, we can carry out the ray tracing procedure in two dimensions for an ellipse with the major axis as $a$ and the minor axis as $b$.

$$R_0 = \frac{ab}{\sqrt{b^2\cos^2\theta + a^2\sin^2\theta}} \quad (39)$$

$$\frac{\partial R_0}{\partial \theta} = \frac{\sin\theta\cos\theta(b^2 - a^2)ab}{(b^2\cos^2\theta + a^2\sin^2\theta)^{3/2}} \quad (40)$$

$$\frac{\partial^2 R_0}{\partial \theta^2} = (b^2 - a^2)ab\left[\frac{\cos 2\theta}{(b^2\cos^2\theta + a^2\sin^2\theta)^{3/2}} + \frac{3}{4}\frac{(b^2-a^2)\sin^2 2\theta}{(b^2\cos^2\theta + a^2\sin^2\theta)^{5/2}}\right] \quad (41)$$

Figure 2 shows the path of a plane wave with the incident wave vector parallel to the Z-axis. The cloaking effect is obvious from the path of the waves. Since it is a symmetric structure, we have generated waves on one side only.

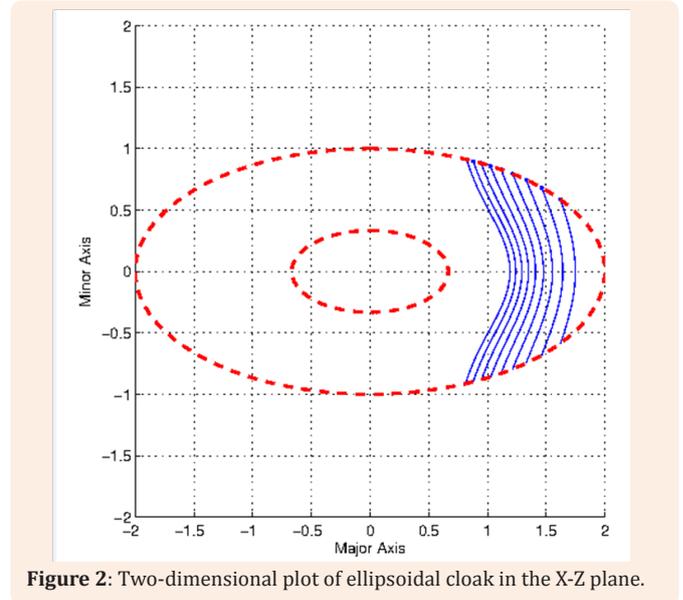

**Figure 2**: Two-dimensional plot of ellipsoidal cloak in the X-Z plane.

## Conclusion

A generalised transformation technique for the design of an arbitrarily shaped cloak in three dimensions has been demonstrated. A procedure for calculating the path of a ray through it is described. In order to demonstrate the validity of the algorithm, it has been applied to an axisymmetric ellipsoid cloak. In the existing algorithms, there is a need for an explicit calculation of material properties ($\mu$ and $\varepsilon$ tensors) and the determinant of the refractive index for the formulation of the Hamiltonian. The algorithm explained in this paper obviates these requirements. Also, the algorithms reported in the literature, for the design of an arbitrarily shaped cloak in three dimensions, have not been verified using the ray tracing technique. Thus there are two distinct contributions of this research work. First, the freedom from the necessity of calculating material properties explicitly and second, the verification of the algorithm using ray tracing. This technique can be applied to any arbitrary surface for which the function $R_0(\theta, \varphi)$ is single valued on it.

## Acknowledgement

None.

## Conflict of Interest

Author declares there is no conflict of interest.